\documentclass[aip,jap,10pt,superscriptaddress,twocolumn,reprint]{revtex4-1}

\usepackage{graphicx}
\usepackage{bpchem}
\usepackage{amssymb}
\usepackage{color}

\begin{document}

\title{First-order ferromagnetic transition in single-crystalline \BPChem{(Mn,Fe)\_{2}(P,Si)}}

\author{H. Yibole} 
 \email{Yibole@tudelft.nl}
 \affiliation{FAME, TUDelft, Mekelweg 15, 2629JB Delft, The Netherlands} 
\author{F. Guillou}
\affiliation{FAME, TUDelft, Mekelweg 15, 2629JB Delft, The Netherlands}
\affiliation{European Synchrotron Radiation Facility, 71 Avenue des Martyrs CS40220, F-38043 Grenoble Cedex 09, France}
\author{Y. K. Huang}
\affiliation{Van der Waals-Zeeman Instituut, Universiteit van Amsterdam, Science Park 904, 1098XH Amsterdam, The Netherlands}
\author{G. R. Blake}
\affiliation{Zernike Institute for Advanced Materials, 9712CP Groningen, The Netherlands}
\author{A. J. E. Lefering}
\affiliation{FAME, TUDelft, Mekelweg 15, 2629JB Delft, The Netherlands}
\author{N.H. van Dijk}
\affiliation{FAME, TUDelft, Mekelweg 15, 2629JB Delft, The Netherlands}
\author{E. Br\"{u}ck}
\affiliation{FAME, TUDelft, Mekelweg 15, 2629JB Delft, The Netherlands}
\date{\today}

\begin{abstract}
\BPChem{(Mn,Fe)\_{2}(P,Si)} single crystals have been successfully grown by flux method. Single crystal diffraction demonstrates that \BPChem{Mn\_{0.83}Fe\_{1.17}P\_{0.72}Si\_{0.28}} crystallizes in a hexagonal crystal structure (space group \textit{P}$\bar{6}$2\textit{m}) at both 100 and 280~K, in the ferromagnetic and paramagnetic states, respectively. The magnetization measurements show that the crystals display a first-order ferromagnetic phase transition at their Curie temperature (\BPChem{\textit{T}\_{C}}). The preferred magnetization direction is along the \textit{c} axis. A weak magnetic anisotropy of \BPChem{\textit{K}\_{1}} $= 0.25\times{10^6}$~$J/m^3$ and \BPChem{\textit{K}\_{2}} $= 0.19\times{10^6}$~$J/m^3$ is found at 5~K. These values indicate a soft magnetic behaviour favourable for magnetic refrigeration. A series of discontinuous magnetization jumps is observed far below \BPChem{\textit{T}\_{C}} by increasing the field at a constant temperature. These magnetization jumps are irreversible, occur spontaneously at constant temperature and magnetic field, but can be restored by cycling across the first-order phase transition.

\end{abstract}
\maketitle

Only a few ferromagnets are known to exhibit a first-order magnetic transition (FOMT) at their \BPChem{\textit{T}\_{C}}. Among those, the \BPChem{Fe\_{2}P}-based magnetocaloric materials have a rich and long history that goes back to the 1960s~\cite{RUN59, FRU69}. Two main dilemmas arose since the very first studies on the binary \BPChem{Fe\_{2}P} compound~\cite{FUJ77, LUN78}. The first point dealt with the nature of the ferromagnetic transition, the exact \BPChem{\textit{T}\_{C}} and the order of the transition. The second point concerned the particularly large magnetocrystalline anisotropy for a 3\textit{d} transition metals compound. A great step towards a better understanding of this materials was made thanks to the synthesis of high purity \BPChem{Fe\_{2}P} single crystals. It is now well accepted that \BPChem{Fe\_{2}P} is a ferromagnet with an easy direction along the \textit{c} axis and it shows a first-order ferromagnetic transition at \BPChem{\textit{T}\_{C}} $\approx$ 216~K. A rare peculiarity is however that the crystal symmetry does not change across the phase transition as the space group remains $P\bar{6}2m$. Examples for this kind of iso-structural first-order transition are scarce, and even more when coupled with a change in magnetic order. 

The manganese alloys derived from \BPChem{Fe\_{2}P} are currently attracting great interest as they display a giant magnetocaloric effect associated with a first-order magneto-elastic transition. This is of interest for many potential applications including magnetic refrigeration. Compared to the parent \BPChem{Fe\_{2}P}, the \BPChem{(Mn,Fe)\_{2}(P,X)} (X = As, Ge, Si) materials are even more exceptional as they display a FOMT one order of magnitude more intense than \BPChem{Fe\_{2}P} and their properties are relatively easy to be controlled by chemical substitutions~\cite{TEG02, TRU09, DUN11, YIB14, GUI14}. 

In spite of extensive experimental data on polycrystalline \BPChem{(Mn,Fe)\_{2}(P,Si)}, the understanding of the magnetic anisotropy and the field-induced metamagnetic transition in these materials is still limited. In particular, as the hexagonal structure is prone to show an anisotropy in the physical properties, single crystals are required to obtain more detailed information. In this study, we report the crystal structure and magnetic properties of high purity single-crystalline \BPChem{(Mn,Fe)\_{2}(P,Si)}. This allows one to supplement previous observations made on polycrystalline samples. In addition, a new magnetic behaviour related to the magnetization process at the ferro-to-paramagnetic FOMT is observed.
\begin{figure}
     \centering
     \includegraphics[trim = 0mm 0mm 0mm 0mm, clip,width=0.45\textwidth]{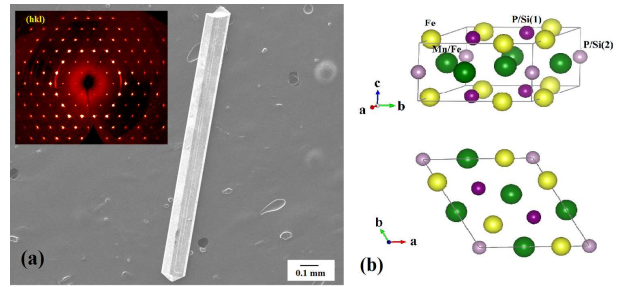}
     \caption{(a) SEM image of as-grown \BPChem{Mn\_{0.83}Fe\_{1.17}P\_{0.72}Si\_{0.28}} single crystals. The inset shows the experimental ($h$ $k$ 0) reciprocal lattice plane at 100~K. (b) Crystal structure of \BPChem{Mn\_{0.83}Fe\_{1.17}P\_{0.72}Si\_{0.28}}.}
     \label{fig1}
\end{figure}

Single crystals of \BPChem{Mn\_{0.83}Fe\_{1.17}P\_{0.72}Si\_{0.28}} were grown by the flux method with tin as metallic flux. Numerous tests have been performed to obtain the optimal conditions ensuring an appropriate silicon content. High-purity Mn (99.9\%), Fe(99.9\%), P(99.9\%) and Si (99.999\%) were used as starting materials. Two approaches have been used for the synthesis of materials before the growth of crystals. For the first method, the starting materials of Mn, Fe, P and Si with nominal composition of \BPChem{Mn\_{0.8}Fe\_{1.2}P\_{0.45}Si\_{0.65}} are mixed with high-purity Sn and then arc-melted under Ar atmosphere in a water-cooled copper crucible. The resulting ingot is then sealed in quartz ampoules in an Ar atmosphere of 200 mbar. The charge to flux ratio was 1:20~wt\%. In the second method, off-stoichiometric polycrystalline samples \BPChem{Mn\_{0.8}Fe\_{1.2}P\_{0.45}Si\_{0.65}} were prepared by high-energy ball milling and annealing as described in~\cite{DUN11}. The pre-synthesized polycrystalline sample was then mixed with Sn and sealed in quartz ampoules. The sealed ampules prepared by the two above methods were placed in a vertical furnace. The thermal treatment for the crystal growth is the following: the temperature is raised to 1473~K in 6.5~h, maintained at this temperature for 100~hours, and then cooled at a rate of 3~K/h to 700~K at which the excess of tin was removed. When necessary, the remaining flux was removed by etching with diluted hydrochloric acid. The two crystals presented in this letter originate from the first method only.  

Microanalyses were performed using JEOL JSM-7500F scanning electron microscope (SEM) equipped with an energy dispersive spectrometer (EDS). The chemical composition was determined with EDS by probing several locations on each crystal. X-ray single crystal diffraction was collected at different temperatures in zero field using a Bruker AXS Kappa APEX II diffractometer equipped with graphite-monochromated Mo K$\alpha$ radiation ($\lambda$ = 0.71073~$\r{A}$). The crystal structure was refined using full-matrix least-squares against $F^2$ (SHELXL-97). 

Magnetic measurements were carried out in a magnetometer (Quantum Design MPMS 5XL) equipped with a reciprocating sample option. Care has been taken in the choice of appropriate ranges for the magnetometer. The mass of the needle single crystal was determined by estimating its volume under SEM and using the density of 6.5~$g/cm^3$ obtained from single crystal diffraction. An uncertainty of 10\% is acknowledged  on the single crystal mass, and thus for the magnetization. For measurements perpendicular to the long axis, the magnetic data are corrected for demagnetizing field. 
\begin{table}
\setlength{\tabcolsep}{6pt}
\caption{\label{table1}Lattice parameters and structure refinement of \BPChem{Mn\_{0.83}Fe\_{1.17}P\_{0.72}Si\_{0.28}} from single crystal diffraction. Atomic positions: 3$g$ ($x_{1}$, 0, 1/2); 3$f$ ($x_{2}$, 0, 0); 2$c$ (1/3, 2/3, 0) and 1$b$ (0, 0, 1/2).}
\begin{ruledtabular}
\begin{tabular}{cccc}
\multicolumn{2}{c}{Temperature (K)}                           & 100                                                         & 280                \\ \hline
\multicolumn{2}{c}{$a$ ($\r{A}$)}                             & 6.0838(9)                                                   & 5.997(6)           \\ 
\multicolumn{2}{c}{$c$ ($\r{A}$)}                             & 3.3556(5)                                                   & 3.484(3)           \\ 
\multicolumn{2}{c}{Volume ($\r{A}^{3}$)}                      & 107.56(4)                                                   & 108.5(2)           \\ 
\multicolumn{2}{c}{$x_{1}$}                                   & 0.5936(2)                                                   & 0.5904(2)           \\ 
\multicolumn{2}{c}{$x_{2}$}                                   & 0.2563(2)                                                   & 0.2548(2)           \\ 
\multicolumn{2}{c}{Collected reflections}                     & 180                                                         & 182                \\ 
\multicolumn{2}{c}{Final R index}                             & 0.0242                                                      & 0.0243             \\ 
\multicolumn{2}{c}{Goodness-of-fit on $F^{2}$}                & 1.096                                                       & 1.050              \\ 
\end{tabular}
\end{ruledtabular}
\end{table}

In \BPChem{(Mn,Fe)\_{2}(P,Si)} materials, small deviations in the Mn/Fe and P/Si ratios can lead to very different crystal and magnetic properties. Due to the poor solubility of the silicon in molten tin, the exact final compositions of the crystals are difficult to predict. We noticed that by starting from nominal compositions with relatively low silicon content (Si = 1/3), the resulting crystals \BPChem{Mn\_{0.96}Fe\_{0.94}P\_{0.8}Si\_{0.2}} belong to the boundary zone of the phase diagram between orthorhombic (\BPChem{Co\_{2}P}-type) and hexagonal (\BPChem{Fe\_{2}P}-type) structures. This range corresponds to an antiferromagnetic order with a \BPChem{\textit{T}\_{N}} $\sim$ 120~K. As we aim to study the ferromagnetic transition, in the rest of this letter we will focus our attention on the crystals specifically grown to present a larger Si content.
\begin{figure}
     \centering
     \includegraphics[trim = 2mm 0mm 0mm 2mm, clip,width=0.45\textwidth]{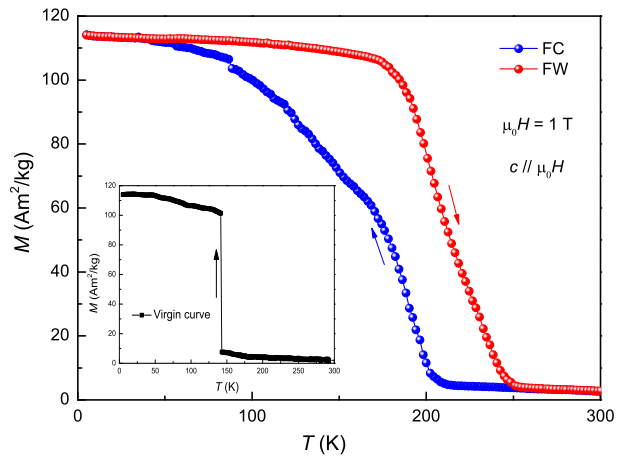}
     \caption{Temperature dependence of the magnetization for \BPChem{Mn\_{0.83}Fe\_{1.17}P\_{0.72}Si\_{0.28}}.  The inset shows the first cooling curve (virgin curve) in similar conditions.}
     \label{fig2}
\end{figure}

The as-grown crystals have a well-formed prismatic shape with average dimensions of $\sim$ 0.15$\times0.15$$\times1.5$ $mm^3$. The surfaces of these crystals are regular and homogeneous. Figure \ref{fig1}(a) depicts a typical example. The chemical formula for this crystal is \BPChem{Mn\_{0.83}Fe\_{1.17}P\_{0.72}Si\_{0.28}}. Single crystal diffraction has been carried out at 280~K (paramagnetic phase) and 100~K (ferromagnetic phase). The inset of Figure \ref{fig1}(a) shows a typical single crystal diffraction pattern. The single crystal diffraction data can be fitted with the \BPChem{Fe\_{2}P} structure, as expected for this range of composition. The structure refinement is carried out on 180 reflexions having an intensity I $\textgreater$ 2$\sigma$ and yield to very satisfactory refinement figure of merits (R $\textless$ 0.025). This demonstrates that the crystals are single phase and of appreciable quality. The lattice parameters for the ferromagnetic and paramagnetic states are listed in Table \ref{table1}, and the unit cell is represented in Figure \ref{fig1}(b). It should be stressed that these values stand for a crystal cycled across the phase transition. The essential result is that these diffraction data confirm the magneto-elastic character of the first-order transition. Across the FOMT, the space group remains $P\bar{6}2m$ and no supplementary reflections are observed. Crossing the ferromagnetic transition upon heating however results in an expansion of the $c$ axis and contraction of the $a$ axis, giving an increase in the $c$/$a$ ratio of + 5.3~\%. As shown in Figure \ref{fig1}(b), the Mn atoms are set in the pyramidal 3$g$ positions and the Fe atoms in both the tetrahedron 3$f$ and the remaining 3$g$ sites. The coordinates of these positions do not evolve across the transition. The P and Si atoms are randomly dispersed between the 2$c$ and 1$b$ sites. These results  confirm the previous report on polycrystals~\cite{DUN11}.

Figure \ref{fig2} presents the magnetization of \BPChem{Mn\_{0.83}Fe\_{1.17}P\_{0.72}Si\_{0.28}} single crystal as a function of temperature for magnetic field of 1~T. The inset shows the first cooling across the FOMT on which an extremely sharp increase in magnetization is observed, as most of the magnetization jump develops in a 1 K temperature increment. The \BPChem{\textit{T}\_{C}} appears lower during this first cooling, a phenomenon often referred to as virgin effect~\cite{BRU08}. The subsequent heating and cooling curves present a much broader transition at \BPChem{\textit{T}\_{C}} = 210 and 170~K for heating and cooling, respectively. Such a large thermal hysteresis of 40 K demonstrates the first-order character of the magneto-elastic phase transition. It is interesting to note that the FOMT in single crystal requires a temperature range as large as in polycrystalline materials to take place. On the cooling curve of Figure \ref{fig2}, the magnetization continues to grow even below 100~K. In many systems, single crystals have shown better magnetocaloric properties than in polycrystalline samples, here no strong improvement is expected.
\begin{figure}
     \centering
     \includegraphics[trim = 20mm 10mm 0mm 20mm, clip,width=0.52\textwidth]{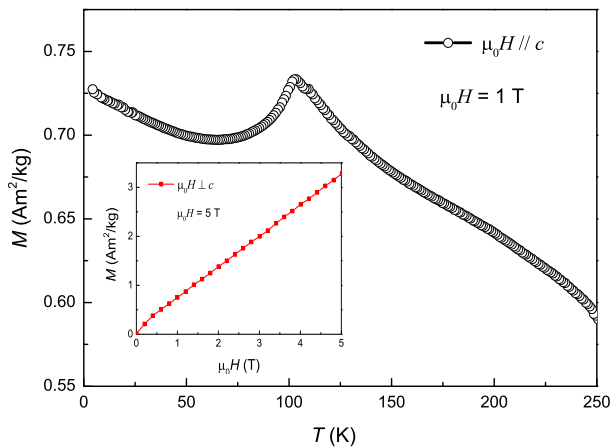}
     \caption{Temperature dependence of the magnetization for \BPChem{Mn\_{0.96}Fe\_{0.94}P\_{0.8}Si\_{0.2}} single crystal. The inset shows the field dependence of the magnetization at 5~K.}
     \label{fig3}
\end{figure}

For comparison, the temperature dependence of the magnetization for \BPChem{Mn\_{0.96}Fe\_{0.94}P\_{0.8}Si\_{0.2}} single crystal (with a lower silicon content) is shown in Figure \ref{fig3}. The low magnetization values below 50~K and the maximum at 105~K indicate an antiferromagnetic ground-state. In the inset of Figure \ref{fig3}, the magnetizations shows a linear dependence on the magnetic field. This illustrates the complexity of the phase diagram of \BPChem{(Mn,Fe)\_{2}(P,Si)} at lower Si content. In particular, the magnetic structure is extremely sensitive to the sample composition.
\begin{figure}
     \centering
     \includegraphics[trim = 20mm 10mm 0mm 15mm, clip,width=0.52\textwidth]{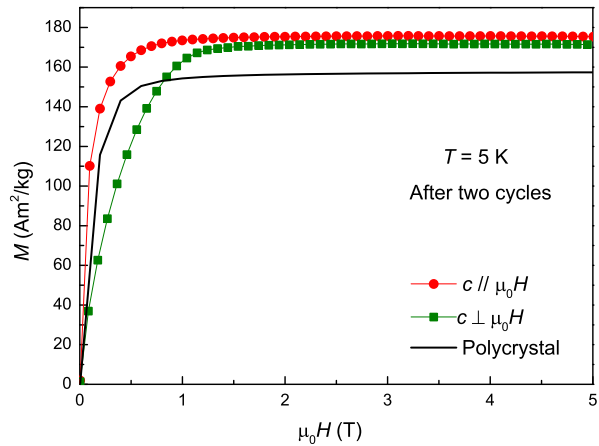}
     \caption{Field dependence of the magnetization for single-crystalline and polycrystal \BPChem{Mn\_{0.83}Fe\_{1.17}P\_{0.72}Si\_{0.28}}.}
     \label{fig4}
\end{figure}

The magnetic anisotropy of the \BPChem{Mn\_{0.83}Fe\_{1.17}P\_{0.72}Si\_{0.28}} crystal (after field cycling at 5~T) has been investigated by measuring the isothermal magnetization along the $c$ axis and in the basal ($a$,$b$) plane. For the field direction perpendicular to the long axis of the crystal, the magnetic field is corrected by using a demagnetization factor of $N$ = 1/2. The results are shown in Figure \ref{fig4} and compared with the magnetization curve for a polycrystalline sample of identical nominal composition. For a magnetic field parallel to the crystallographic $c$ axis, the magnetization increases very rapidly and saturates at fields lower than 1~T. On the other hand, a significantly larger field is needed to reach saturation in the field direction perpendicular to the $c$ axis. The saturation magnetization is \BPChem{4.5~{$\mu$}\_{B}} at 5~K, which appears to be slightly higher than that of the polycrystalline sample. This small discrepancy of the saturation magnetizations between single crystals and polycrystalline sample can be explained by a small amount of impurity phases with lower magnetization in the latter. The saturation magnetization of the single crystal is well compatible with previous reports for closely related compositions on the basis of bulk magnetometry, neutron diffraction~\cite{OU13}. Here, the magnetic anisotropy has been quantitatively estimated. As usual for a hexagonal system, we took only the first- and second- order anisotropy constants into consideration, so that the magnetocrystalline anisotropy energy can be express as $E$ $\approx$ \BPChem{\textit{K}\_{1}}$\sin^2 \theta$ + \BPChem{\textit{K}\_{2}}$\sin^4 \theta$, where $\theta$ is the polar angle between the $c$ axis and the magnetization vector. The magnetocrystalline anisotropy constants are deduced using the method developed by Sucksmith and Thompson~\cite{SUC54}. The results are \BPChem{\textit{K}\_{1}} $= 0.25\times{10^6}$ $J/m^3$ and \BPChem{\textit{K}\_{2}} $= 0.19\times{10^6}$ $J/m^3$ at 5 K. This direct and quantitative determination of the magnetocrystalline anisotropy shows that \BPChem{\textit{K}\_{1}} predominates and is positive. So, the preferred magnetization direction in \BPChem{Mn\_{0.83}Fe\_{1.17}P\_{0.72}Si\_{0.28}} crystal is along the $c$ axis. This is consistent with the results from earlier neutron diffraction study stating that the easy magnetization direction in \BPChem{MnFe(P\_{1-x}Si\_{x})} compounds lays close to the $c$ axis, with an angle $\theta$ = $29\,^{\circ}$ with [001] for x = 0.34~\cite{OU13}. These values of anisotropy constants are one order of magnitude lower than that of the parent compound \BPChem{Fe\_{2}P}~\cite{FUJ77, CAR13}. It should be stressed that this relatively limited magneto-crystalline anisotropy in \BPChem{(Mn,Fe)\_{2}(P,Si)} is an advantage for large scale application of magnetic refrigerants.
\begin{figure}
     \centering
     \includegraphics[trim = 25mm 80mm 15mm 20mm, clip,width=0.55\textwidth]{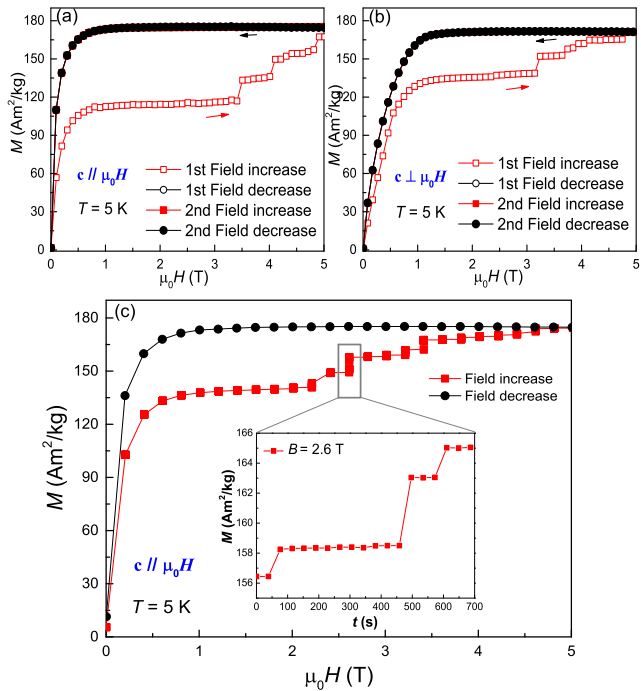}
     \caption{ (a) and (b) Field dependence of the magnetization for single-crystalline \BPChem{Mn\_{0.83}Fe\_{1.17}P\_{0.72}Si\_{0.28}} at 5 K after ZFC for two principal directions. (c) Magnetization as a function of field measured along the $c$ axis for single-crystalline \BPChem{Mn\_{0.83}Fe\_{1.17}P\_{0.72}Si\_{0.28}} at $T$ = 5~K (after ZFC). The inset shows the enlargement of the magnetic relaxation for the magnetic field of 2.6~T.}
     \label{fig5}
\end{figure}

Figure \ref{fig5} shows the initial magnetization curves taken prior to the data presented in Figure \ref{fig4} after a zero-field cooling (ZFC). In Figure \ref{fig5}(a), for the first field increase $M$($H$), the most striking feature occurring during the magnetization process is well noticeable. A plateau of almost constant magnetization (113~\BPChem{Am\^{2}kg\^{-1}}) is observed until around 3~T, in agreement with the $M(T)$ measurement shown in Figure \ref{fig2}. Above 3~T, a series of step-like magnetization jumps occurs. These unusual magnetization jumps also occur along the direction perpendicular to the $c$ axis, as shown in Figure \ref{fig5}(b). It can be noticed that, even though a large magnetic field is systematically required, the magnetization jumps do not occur at regular field intervals. The magnetization jumps need to overcome a certain magnetic field threshold to fully develop. In other respects, jumps of opposite direction were never observed during the demagnetization from 5~T to zero magnetic field. In addition, they are irreversible in the sense that they are not observed during a subsequent magnetization/demagnetization cycle (at constant temperature), as shown in Figure \ref{fig5}(a) and (b). It is important to note that the first field-increase $M(H)$ curves in Figure \ref{fig5} (a) and (b) were measured with a ZFC preparation in each case, i.e., there has been a thermal cycling across the phase transition. As a matter of fact, the present magnetization jumps observed in single crystals do \emph{not} correspond to the usual virgin effect reported till now in polycrystalline samples, as this later disappears after one thermal cycling. Here, the magnetization jumps can be recovered after resetting the materials by a cycling across the FOMT. However, after many thermal cycles (n $\textgreater$ 5), or when the cooling is carried out in different magnetic field ($H\neq{0}$), we observed that the field required for the magnetic jumps evolves (in random directions), so that this phenomenon remains to some extend sensitive to the thermal and magnetic histories.

To check whether these magnetization jumps are related to a dynamical effect during the magnetization process, we measured the magnetic relaxation at 5~K for the \BPChem{Mn\_{0.83}Fe\_{1.17}P\_{0.72}Si\_{0.28}} crystal, at each magnetic field during a magnetization from zero to 5~T (after a ZFC). As shown in Figure \ref{fig5}(c), the magnetization jumps occur spontaneously at constant temperature and magnetic field. In the inset of Figure \ref{fig5}(c), several magnetization plateaus separated by the magnetization jumps can be distinguished during a single relaxation curve. The results show a strong time-dependent effect and indicate the existence of several non-equilibrium metastable states. The time scale of these magnetization jumps (several hundreds of second after applying the magnetic field) and the magnetic field range ($\textgreater$ 2.5~T) indicate that we are dealing with a phenomenon related to the phase transition, and not with more generic magnetization processes such as magnetic field screening by eddy current (much smaller time scale) or domain wall movements (relevant at lower magnetic field).

Such a step-like magnetic behaviour is a new phenomenon in \BPChem{(Mn,Fe)\_{2}(P,Si)} compounds. We believe that two scenarios (or a combination of both) could possibly explain our results: (i) an underlying antiferromagnetic order, and (ii) a dynamical phase separation phenomena. The first interpretation is specific to \BPChem{(Mn,Fe)\_{2}(P,Si)} materials. In \BPChem{FeMnP\_{0.75}Si\_{0.25}} polycrystalline samples, it has recently been highlighted by neutron diffraction experiments and first principles calculations that ferromagnetism and a metastable antiferromagnetic phase can coexist below \BPChem{\textit{T}\_{C}}~\cite{HOG15, LI14}. These observations are made on the Si poor edge of the phase diagrams, from which our single crystals are relatively close. In this context, the magnetization jumps would correspond to partial field induced antifero-to-ferromagnetic transition. The second scenario is more general to FOMT. A similar behaviour of successive magnetic discontinuities have been also observed in a wide variety of materials as long as it is associated to a FOMT~\cite{MAH02, HAR04, LYU08, ROY05, GIV83}. Depending on the system, different interpretations have been proposed. The most frequent explanation is the so-called martensitic scenario~\cite{MAH02, HAR04}. It relies on the fact that the two magnetic phases at play correspond to two incompatibles crystallographic phases (either different lattice symmetry or different cell parameters), so that the growth of one phase is done at the expense of the other and will result in interfacial constraints. Upon cooling across the FOMT, the interfacial strains created at the ferromagnetic/paramagnetic interface will result in an arrest of the transition and a fraction of the crystal will be blocked into metastable paramagnetic states far below the equilibrium \BPChem{\textit{T}\_{C}}. At low temperatures, as the magnetic field is increased, the driving force aligning the spins becomes strong enough to overcome the energy barriers from the interfacial strains, so the ferromagnetic phase experiences a burst-like growth and leads to a magnetization jump. Both interpretations for the magnetic jumps are plausible. Note that single crystal diffraction is of little help, as the diffractograms at 100 K are clearly single phase. Whereas in both scenarios, either the metastable antiferromagnetic phase or the untransformed paramagnetic phase show a similar $c$/$a$ ratio, significantly different from the ferromagnetic state, so that their contribution would have been noticed. A spatial resolution such as that provided by tomography techniques is required to investigate the distribution in the crystal of the different phases.

To conclude, \BPChem{(Mn,Fe)\_{2}(P,Si)} single crystals presenting a first-order ferro-to-paramagnetic transition have been grown for the first time. The studies of their structural and magnetic properties bring support to the previous reports on polycrystalline materials. A weak magnetic anisotropy is found at 5~K indicating a soft magnetic behaviour favourable for magnetic refrigeration. The magnetization process toward the ferromagnetic state turns out to be more complex in single crystals than for polycrystalline samples. Series of discontinuous magnetization jumps can be observed far below the Curie temperature. These jumps are irreversible, but can be restored by resetting the crystal across the FOMT. 

The authors would like to thank Bert Zwart for his help in sample preparation. This work is financially supported by the Foundation for Fundamental Research on Matter (FOM) (The Netherlands) and BASF New Business.

\providecommand{\noopsort}[1]{}\providecommand{\singleletter}[1]{#1}%


\begin{thebibliography}{20}%
\makeatletter
\providecommand \@ifxundefined [1]{%
 \@ifx{#1\undefined}
}%
\providecommand \@ifnum [1]{%
 \ifnum #1\expandafter \@firstoftwo
 \else \expandafter \@secondoftwo
 \fi
}%
\providecommand \@ifx [1]{%
 \ifx #1\expandafter \@firstoftwo
 \else \expandafter \@secondoftwo
 \fi
}%
\providecommand \natexlab [1]{#1}%
\providecommand \enquote  [1]{``#1''}%
\providecommand \bibnamefont  [1]{#1}%
\providecommand \bibfnamefont [1]{#1}%
\providecommand \citenamefont [1]{#1}%
\providecommand \href@noop [0]{\@secondoftwo}%
\providecommand \href [0]{\begingroup \@sanitize@url \@href}%
\providecommand \@href[1]{\@@startlink{#1}\@@href}%
\providecommand \@@href[1]{\endgroup#1\@@endlink}%
\providecommand \@sanitize@url [0]{\catcode `\\12\catcode `\$12\catcode
  `\&12\catcode `\#12\catcode `\^12\catcode `\_12\catcode `\%12\relax}%
\providecommand \@@startlink[1]{}%
\providecommand \@@endlink[0]{}%
\providecommand \url  [0]{\begingroup\@sanitize@url \@url }%
\providecommand \@url [1]{\endgroup\@href {#1}{\urlprefix }}%
\providecommand \urlprefix  [0]{URL }%
\providecommand \Eprint [0]{\href }%
\providecommand \doibase [0]{http://dx.doi.org/}%
\providecommand \selectlanguage [0]{\@gobble}%
\providecommand \bibinfo  [0]{\@secondoftwo}%
\providecommand \bibfield  [0]{\@secondoftwo}%
\providecommand \translation [1]{[#1]}%
\providecommand \BibitemOpen [0]{}%
\providecommand \bibitemStop [0]{}%
\providecommand \bibitemNoStop [0]{.\EOS\space}%
\providecommand \EOS [0]{\spacefactor3000\relax}%
\providecommand \BibitemShut  [1]{\csname bibitem#1\endcsname}%
\let\auto@bib@innerbib\@empty
\bibitem [{\citenamefont {Rundqvist}\ and\ \citenamefont
  {Jellinek}(1959)}]{RUN59}%
  \BibitemOpen
  \bibfield  {author} {\bibinfo {author} {\bibfnamefont {S.}~\bibnamefont
  {Rundqvist}}\ and\ \bibinfo {author} {\bibfnamefont {F.}~\bibnamefont
  {Jellinek}},\ }\href {http://dx.doi.org/10.3891/acta.chem.scand.13-0425}
  {}Vol.~\bibinfo {volume} {13}\ (\bibinfo {year} {1959})\ p.\ \bibinfo {pages}
  {425}\BibitemShut {NoStop}%
\bibitem [{\citenamefont {R.~Fruchart}\ and\ \citenamefont
  {Senateur}(1969)}]{FRU69}%
  \BibitemOpen
  \bibfield  {author} {\bibinfo {author} {\bibfnamefont {A.~R.}\ \bibnamefont
  {R.~Fruchart}}\ and\ \bibinfo {author} {\bibfnamefont {J.~P.}\ \bibnamefont
  {Senateur}},\ }\href {http://dx.doi.org/10.1063/1.1657617} {\bibfield
  {journal} {\bibinfo  {journal} {J. Appl. Phys.}\ }\textbf {\bibinfo {volume}
  {40}},\ \bibinfo {pages} {1250} (\bibinfo {year} {1969})}\BibitemShut
  {NoStop}%
\bibitem [{\citenamefont {Fujii}\ \emph {et~al.}(1977)\citenamefont {Fujii},
  \citenamefont {Hōkabe}, \citenamefont {Kamigaichi},\ and\ \citenamefont
  {Okamoto}}]{FUJ77}%
  \BibitemOpen
  \bibfield  {author} {\bibinfo {author} {\bibfnamefont {H.}~\bibnamefont
  {Fujii}}, \bibinfo {author} {\bibfnamefont {T.}~\bibnamefont {Hōkabe}},
  \bibinfo {author} {\bibfnamefont {T.}~\bibnamefont {Kamigaichi}}, \ and\
  \bibinfo {author} {\bibfnamefont {T.}~\bibnamefont {Okamoto}},\ }\href
  {\doibase 10.1143/JPSJ.43.41} {\bibfield  {journal} {\bibinfo  {journal} {J.
  Phys. Soc. Jpn.}\ }\textbf {\bibinfo {volume} {43}},\ \bibinfo {pages} {41}
  (\bibinfo {year} {1977})}\BibitemShut {NoStop}%
\bibitem [{\citenamefont {Lundgren}\ \emph {et~al.}(1978)\citenamefont
  {Lundgren}, \citenamefont {Tarmohamed}, \citenamefont {Beckman},
  \citenamefont {Carlsson},\ and\ \citenamefont {Rundqvist}}]{LUN78}%
  \BibitemOpen
  \bibfield  {author} {\bibinfo {author} {\bibfnamefont {L.}~\bibnamefont
  {Lundgren}}, \bibinfo {author} {\bibfnamefont {G.}~\bibnamefont
  {Tarmohamed}}, \bibinfo {author} {\bibfnamefont {O.}~\bibnamefont {Beckman}},
  \bibinfo {author} {\bibfnamefont {B.}~\bibnamefont {Carlsson}}, \ and\
  \bibinfo {author} {\bibfnamefont {S.}~\bibnamefont {Rundqvist}},\ }\href
  {\doibase 10.1088/0031-8949/17/1/008} {\bibfield  {journal} {\bibinfo
  {journal} {Phys. Scripta}\ }\textbf {\bibinfo {volume} {17}},\ \bibinfo
  {pages} {39} (\bibinfo {year} {1978})}\BibitemShut {NoStop}%
\bibitem [{\citenamefont {Tegus}\ \emph {et~al.}(2002)\citenamefont {Tegus},
  \citenamefont {Br\"uck}, \citenamefont {Buschow},\ and\ \citenamefont
  {de~Boer}}]{TEG02}%
  \BibitemOpen
  \bibfield  {author} {\bibinfo {author} {\bibfnamefont {O.}~\bibnamefont
  {Tegus}}, \bibinfo {author} {\bibfnamefont {E.}~\bibnamefont {Br\"uck}},
  \bibinfo {author} {\bibfnamefont {K.~H.~J.}\ \bibnamefont {Buschow}}, \ and\
  \bibinfo {author} {\bibfnamefont {F.~R.}\ \bibnamefont {de~Boer}},\ }\href
  {\doibase 10.1038/415150a} {\bibfield  {journal} {\bibinfo  {journal}
  {Nature}\ }\textbf {\bibinfo {volume} {415}},\ \bibinfo {pages} {150}
  (\bibinfo {year} {2002})}\BibitemShut {NoStop}%
\bibitem [{\citenamefont {Trung}\ \emph {et~al.}(2009)\citenamefont {Trung},
  \citenamefont {Ou}, \citenamefont {Gortenmulder}, \citenamefont {Tegus},
  \citenamefont {Buschow},\ and\ \citenamefont {Br\"uck}}]{TRU09}%
  \BibitemOpen
  \bibfield  {author} {\bibinfo {author} {\bibfnamefont {N.~T.}\ \bibnamefont
  {Trung}}, \bibinfo {author} {\bibfnamefont {Z.~Q.}\ \bibnamefont {Ou}},
  \bibinfo {author} {\bibfnamefont {T.~J.}\ \bibnamefont {Gortenmulder}},
  \bibinfo {author} {\bibfnamefont {O.}~\bibnamefont {Tegus}}, \bibinfo
  {author} {\bibfnamefont {K.~H.~J.}\ \bibnamefont {Buschow}}, \ and\ \bibinfo
  {author} {\bibfnamefont {E.}~\bibnamefont {Br\"uck}},\ }\href {\doibase
  http://dx.doi.org/10.1063/1.3095597} {\bibfield  {journal} {\bibinfo
  {journal} {Appl. Phys. Lett.}\ }\textbf {\bibinfo {volume} {94}},\ \bibinfo
  {pages} {102513} (\bibinfo {year} {2009})}\BibitemShut {NoStop}%
\bibitem [{\citenamefont {Dung}\ \emph {et~al.}(2011)\citenamefont {Dung},
  \citenamefont {Ou}, \citenamefont {Caron}, \citenamefont {Zhang},
  \citenamefont {Thanh}, \citenamefont {de~Wijs}, \citenamefont {de~Groot},
  \citenamefont {Buschow},\ and\ \citenamefont {Br\"uck}}]{DUN11}%
  \BibitemOpen
  \bibfield  {author} {\bibinfo {author} {\bibfnamefont {N.~H.}\ \bibnamefont
  {Dung}}, \bibinfo {author} {\bibfnamefont {Z.~Q.}\ \bibnamefont {Ou}},
  \bibinfo {author} {\bibfnamefont {L.}~\bibnamefont {Caron}}, \bibinfo
  {author} {\bibfnamefont {L.}~\bibnamefont {Zhang}}, \bibinfo {author}
  {\bibfnamefont {D.~T.~C.}\ \bibnamefont {Thanh}}, \bibinfo {author}
  {\bibfnamefont {G.~A.}\ \bibnamefont {de~Wijs}}, \bibinfo {author}
  {\bibfnamefont {R.~A.}\ \bibnamefont {de~Groot}}, \bibinfo {author}
  {\bibfnamefont {K.~H.~J.}\ \bibnamefont {Buschow}}, \ and\ \bibinfo {author}
  {\bibfnamefont {E.}~\bibnamefont {Br\"uck}},\ }\href {\doibase
  10.1002/aenm.201100252} {\bibfield  {journal} {\bibinfo  {journal} {Adv.
  Energy Mater.}\ }\textbf {\bibinfo {volume} {1}},\ \bibinfo {pages} {1215}
  (\bibinfo {year} {2011})}\BibitemShut {NoStop}%
\bibitem [{\citenamefont {Yibole}\ \emph {et~al.}(2014)\citenamefont {Yibole},
  \citenamefont {Guillou}, \citenamefont {Zhang}, \citenamefont {van Dijk},\
  and\ \citenamefont {Br\"uck}}]{YIB14}%
  \BibitemOpen
  \bibfield  {author} {\bibinfo {author} {\bibfnamefont {H.}~\bibnamefont
  {Yibole}}, \bibinfo {author} {\bibfnamefont {F.}~\bibnamefont {Guillou}},
  \bibinfo {author} {\bibfnamefont {L.}~\bibnamefont {Zhang}}, \bibinfo
  {author} {\bibfnamefont {N.~H.}\ \bibnamefont {van Dijk}}, \ and\ \bibinfo
  {author} {\bibfnamefont {E.}~\bibnamefont {Br\"uck}},\ }\href
  {http://stacks.iop.org/0022-3727/47/i=7/a=075002} {\bibfield  {journal}
  {\bibinfo  {journal} {J. Phys. D: Appl. Phys.}\ }\textbf {\bibinfo {volume}
  {47}},\ \bibinfo {pages} {075002} (\bibinfo {year} {2014})}\BibitemShut
  {NoStop}%
\bibitem [{\citenamefont {Guillou}\ \emph {et~al.}(2014)\citenamefont
  {Guillou}, \citenamefont {Porcari}, \citenamefont {Yibole}, \citenamefont
  {van Dijk},\ and\ \citenamefont {Br\"uck}}]{GUI14}%
  \BibitemOpen
  \bibfield  {author} {\bibinfo {author} {\bibfnamefont {F.}~\bibnamefont
  {Guillou}}, \bibinfo {author} {\bibfnamefont {G.}~\bibnamefont {Porcari}},
  \bibinfo {author} {\bibfnamefont {H.}~\bibnamefont {Yibole}}, \bibinfo
  {author} {\bibfnamefont {N.}~\bibnamefont {van Dijk}}, \ and\ \bibinfo
  {author} {\bibfnamefont {E.}~\bibnamefont {Br\"uck}},\ }\href {\doibase
  10.1002/adma.201304788} {\bibfield  {journal} {\bibinfo  {journal} {Adv.
  Mater.}\ }\textbf {\bibinfo {volume} {26}},\ \bibinfo {pages} {2671}
  (\bibinfo {year} {2014})}\BibitemShut {NoStop}%
\bibitem [{\citenamefont {Br\"uck}\ \emph {et~al.}(2008)\citenamefont
  {Br\"uck}, \citenamefont {Tegus}, \citenamefont {Thanh}, \citenamefont
  {Trung},\ and\ \citenamefont {Buschow}}]{BRU08}%
  \BibitemOpen
  \bibfield  {author} {\bibinfo {author} {\bibfnamefont {E.}~\bibnamefont
  {Br\"uck}}, \bibinfo {author} {\bibfnamefont {O.}~\bibnamefont {Tegus}},
  \bibinfo {author} {\bibfnamefont {D.~C.}\ \bibnamefont {Thanh}}, \bibinfo
  {author} {\bibfnamefont {N.~T.}\ \bibnamefont {Trung}}, \ and\ \bibinfo
  {author} {\bibfnamefont {K.}~\bibnamefont {Buschow}},\ }\href {\doibase
  http://dx.doi.org/10.1016/j.ijrefrig.2007.11.013} {\bibfield  {journal}
  {\bibinfo  {journal} {Int. J. Refrig.}\ }\textbf {\bibinfo {volume} {31}},\
  \bibinfo {pages} {763 } (\bibinfo {year} {2008})}\BibitemShut {NoStop}%
\bibitem [{\citenamefont {Ou}\ \emph {et~al.}(2013)\citenamefont {Ou},
  \citenamefont {Zhang}, \citenamefont {Dung}, \citenamefont {van Eijck},
  \citenamefont {Mulders}, \citenamefont {Avdeev}, \citenamefont {van Dijk},\
  and\ \citenamefont {Br\"uck}}]{OU13}%
  \BibitemOpen
  \bibfield  {author} {\bibinfo {author} {\bibfnamefont {Z.}~\bibnamefont
  {Ou}}, \bibinfo {author} {\bibfnamefont {L.}~\bibnamefont {Zhang}}, \bibinfo
  {author} {\bibfnamefont {N.}~\bibnamefont {Dung}}, \bibinfo {author}
  {\bibfnamefont {L.}~\bibnamefont {van Eijck}}, \bibinfo {author}
  {\bibfnamefont {A.}~\bibnamefont {Mulders}}, \bibinfo {author} {\bibfnamefont
  {M.}~\bibnamefont {Avdeev}}, \bibinfo {author} {\bibfnamefont
  {N.}~\bibnamefont {van Dijk}}, \ and\ \bibinfo {author} {\bibfnamefont
  {E.}~\bibnamefont {Br\"uck}},\ }\href {\doibase
  http://dx.doi.org/10.1016/j.jmmm.2013.03.028} {\bibfield  {journal} {\bibinfo
   {journal} {J. Magn. Magn. Mater.}\ }\textbf {\bibinfo {volume} {340}},\
  \bibinfo {pages} {80 } (\bibinfo {year} {2013})}\BibitemShut {NoStop}%
\bibitem [{\citenamefont {Sucksmith}\ and\ \citenamefont
  {Thompson}(1985)}]{SUC54}%
  \BibitemOpen
  \bibfield  {author} {\bibinfo {author} {\bibfnamefont {W.}~\bibnamefont
  {Sucksmith}}\ and\ \bibinfo {author} {\bibfnamefont {J.~E.}\ \bibnamefont
  {Thompson}},\ }\href {\doibase 10.1098/rspa.1954.0209} {\bibfield  {journal}
  {\bibinfo  {journal} {Proc. Royl. Soc. A}\ }\textbf {\bibinfo {volume}
  {225}},\ \bibinfo {pages} {362} (\bibinfo {year} {1985})}\BibitemShut
  {NoStop}%
\bibitem [{\citenamefont {Caron}\ \emph {et~al.}(2013)\citenamefont {Caron},
  \citenamefont {Hudl}, \citenamefont {H\"oglin}, \citenamefont {Dung},
  \citenamefont {Gomez}, \citenamefont {Sahlberg}, \citenamefont {Br\"uck},
  \citenamefont {Andersson},\ and\ \citenamefont {Nordblad}}]{CAR13}%
  \BibitemOpen
  \bibfield  {author} {\bibinfo {author} {\bibfnamefont {L.}~\bibnamefont
  {Caron}}, \bibinfo {author} {\bibfnamefont {M.}~\bibnamefont {Hudl}},
  \bibinfo {author} {\bibfnamefont {V.}~\bibnamefont {H\"oglin}}, \bibinfo
  {author} {\bibfnamefont {N.~H.}\ \bibnamefont {Dung}}, \bibinfo {author}
  {\bibfnamefont {C.~P.}\ \bibnamefont {Gomez}}, \bibinfo {author}
  {\bibfnamefont {M.}~\bibnamefont {Sahlberg}}, \bibinfo {author}
  {\bibfnamefont {E.}~\bibnamefont {Br\"uck}}, \bibinfo {author} {\bibfnamefont
  {Y.}~\bibnamefont {Andersson}}, \ and\ \bibinfo {author} {\bibfnamefont
  {P.}~\bibnamefont {Nordblad}},\ }\href {\doibase 10.1103/PhysRevB.88.094440}
  {\bibfield  {journal} {\bibinfo  {journal} {Phys. Rev. B}\ }\textbf {\bibinfo
  {volume} {88}},\ \bibinfo {pages} {094440} (\bibinfo {year}
  {2013})}\BibitemShut {NoStop}%
\bibitem [{\citenamefont {H\"oglin}\ \emph {et~al.}(2015)\citenamefont
  {H\"oglin}, \citenamefont {Hudl}, \citenamefont {Caron}, \citenamefont
  {Beran}, \citenamefont {S{\o}rby}, \citenamefont {Nordblad}, \citenamefont
  {Andersson},\ and\ \citenamefont {Sahlberg}}]{HOG15}%
  \BibitemOpen
  \bibfield  {author} {\bibinfo {author} {\bibfnamefont {V.}~\bibnamefont
  {H\"oglin}}, \bibinfo {author} {\bibfnamefont {M.}~\bibnamefont {Hudl}},
  \bibinfo {author} {\bibfnamefont {L.}~\bibnamefont {Caron}}, \bibinfo
  {author} {\bibfnamefont {P.}~\bibnamefont {Beran}}, \bibinfo {author}
  {\bibfnamefont {M.~H.}\ \bibnamefont {S{\o}rby}}, \bibinfo {author}
  {\bibfnamefont {P.}~\bibnamefont {Nordblad}}, \bibinfo {author}
  {\bibfnamefont {Y.}~\bibnamefont {Andersson}}, \ and\ \bibinfo {author}
  {\bibfnamefont {M.}~\bibnamefont {Sahlberg}},\ }\href {\doibase
  http://dx.doi.org/10.1016/j.jssc.2014.10.013} {\bibfield  {journal} {\bibinfo
   {journal} {J. Phys. Chem. Solids}\ }\textbf {\bibinfo {volume} {221}},\
  \bibinfo {pages} {240} (\bibinfo {year} {2015})}\BibitemShut {NoStop}%
\bibitem [{\citenamefont {Li}\ \emph {et~al.}(2014)\citenamefont {Li},
  \citenamefont {Li}, \citenamefont {Sch\"onecker}, \citenamefont {Li},
  \citenamefont {Delczeg-Czirjak}, \citenamefont {Kvashnin}, \citenamefont
  {Eriksson}, \citenamefont {Johansson},\ and\ \citenamefont {Vitos}}]{LI14}%
  \BibitemOpen
  \bibfield  {author} {\bibinfo {author} {\bibfnamefont {G.}~\bibnamefont
  {Li}}, \bibinfo {author} {\bibfnamefont {W.}~\bibnamefont {Li}}, \bibinfo
  {author} {\bibfnamefont {S.}~\bibnamefont {Sch\"onecker}}, \bibinfo {author}
  {\bibfnamefont {X.}~\bibnamefont {Li}}, \bibinfo {author} {\bibfnamefont
  {E.~K.}\ \bibnamefont {Delczeg-Czirjak}}, \bibinfo {author} {\bibfnamefont
  {Y.~O.}\ \bibnamefont {Kvashnin}}, \bibinfo {author} {\bibfnamefont
  {O.}~\bibnamefont {Eriksson}}, \bibinfo {author} {\bibfnamefont
  {B.}~\bibnamefont {Johansson}}, \ and\ \bibinfo {author} {\bibfnamefont
  {L.}~\bibnamefont {Vitos}},\ }\href
  {http://scitation.aip.org/content/aip/journal/apl/105/26/10.1063/1.4905270",
  doi = "http://dx.doi.org/10.1063/1.4905270} {\bibfield  {journal} {\bibinfo
  {journal} {Appl. Phys. Lett.}\ }\textbf {\bibinfo {volume} {105}},\ \bibinfo
  {pages} {262405} (\bibinfo {year} {2014})}\BibitemShut {NoStop}%
\bibitem [{\citenamefont {Mahendiran}\ \emph {et~al.}(2002)\citenamefont
  {Mahendiran}, \citenamefont {Maignan}, \citenamefont {H\'ebert},
  \citenamefont {Martin}, \citenamefont {Hervieu}, \citenamefont {Raveau},
  \citenamefont {Mitchell},\ and\ \citenamefont {Schiffer}}]{MAH02}%
  \BibitemOpen
  \bibfield  {author} {\bibinfo {author} {\bibfnamefont {R.}~\bibnamefont
  {Mahendiran}}, \bibinfo {author} {\bibfnamefont {A.}~\bibnamefont {Maignan}},
  \bibinfo {author} {\bibfnamefont {S.}~\bibnamefont {H\'ebert}}, \bibinfo
  {author} {\bibfnamefont {C.}~\bibnamefont {Martin}}, \bibinfo {author}
  {\bibfnamefont {M.}~\bibnamefont {Hervieu}}, \bibinfo {author} {\bibfnamefont
  {B.}~\bibnamefont {Raveau}}, \bibinfo {author} {\bibfnamefont {J.~F.}\
  \bibnamefont {Mitchell}}, \ and\ \bibinfo {author} {\bibfnamefont
  {P.}~\bibnamefont {Schiffer}},\ }\href {\doibase
  10.1103/PhysRevLett.89.286602} {\bibfield  {journal} {\bibinfo  {journal}
  {Phys. Rev. Lett.}\ }\textbf {\bibinfo {volume} {89}},\ \bibinfo {pages}
  {286602} (\bibinfo {year} {2002})}\BibitemShut {NoStop}%
\bibitem [{\citenamefont {Hardy}\ \emph {et~al.}(2004)\citenamefont {Hardy},
  \citenamefont {Majumdar}, \citenamefont {Crowe}, \citenamefont {Lees},
  \citenamefont {Paul}, \citenamefont {Herv\'e}, \citenamefont {Maignan},
  \citenamefont {H\'ebert}, \citenamefont {Martin}, \citenamefont {Yaicle},
  \citenamefont {Hervieu},\ and\ \citenamefont {Raveau}}]{HAR04}%
  \BibitemOpen
  \bibfield  {author} {\bibinfo {author} {\bibfnamefont {V.}~\bibnamefont
  {Hardy}}, \bibinfo {author} {\bibfnamefont {S.}~\bibnamefont {Majumdar}},
  \bibinfo {author} {\bibfnamefont {S.~J.}\ \bibnamefont {Crowe}}, \bibinfo
  {author} {\bibfnamefont {M.~R.}\ \bibnamefont {Lees}}, \bibinfo {author}
  {\bibfnamefont {D.~M.}\ \bibnamefont {Paul}}, \bibinfo {author}
  {\bibfnamefont {L.}~\bibnamefont {Herv\'e}}, \bibinfo {author} {\bibfnamefont
  {A.}~\bibnamefont {Maignan}}, \bibinfo {author} {\bibfnamefont
  {S.}~\bibnamefont {H\'ebert}}, \bibinfo {author} {\bibfnamefont
  {C.}~\bibnamefont {Martin}}, \bibinfo {author} {\bibfnamefont
  {C.}~\bibnamefont {Yaicle}}, \bibinfo {author} {\bibfnamefont
  {M.}~\bibnamefont {Hervieu}}, \ and\ \bibinfo {author} {\bibfnamefont
  {B.}~\bibnamefont {Raveau}},\ }\href {\doibase 10.1103/PhysRevB.69.020407}
  {\bibfield  {journal} {\bibinfo  {journal} {Phys. Rev. B}\ }\textbf {\bibinfo
  {volume} {69}},\ \bibinfo {pages} {020407} (\bibinfo {year}
  {2004})}\BibitemShut {NoStop}%
\bibitem [{\citenamefont {Lyubina}\ \emph {et~al.}(2008)\citenamefont
  {Lyubina}, \citenamefont {Nenkov}, \citenamefont {Schultz},\ and\
  \citenamefont {Gutfleisch}}]{LYU08}%
  \BibitemOpen
  \bibfield  {author} {\bibinfo {author} {\bibfnamefont {J.}~\bibnamefont
  {Lyubina}}, \bibinfo {author} {\bibfnamefont {K.}~\bibnamefont {Nenkov}},
  \bibinfo {author} {\bibfnamefont {L.}~\bibnamefont {Schultz}}, \ and\
  \bibinfo {author} {\bibfnamefont {O.}~\bibnamefont {Gutfleisch}},\ }\href
  {\doibase 10.1103/PhysRevLett.101.177203} {\bibfield  {journal} {\bibinfo
  {journal} {Phys. Rev. Lett.}\ }\textbf {\bibinfo {volume} {101}},\ \bibinfo
  {pages} {177203} (\bibinfo {year} {2008})}\BibitemShut {NoStop}%
\bibitem [{\citenamefont {Roy}\ \emph {et~al.}(2005)\citenamefont {Roy},
  \citenamefont {Chattopadhyay}, \citenamefont {Chaddah},\ and\ \citenamefont
  {Nigam}}]{ROY05}%
  \BibitemOpen
  \bibfield  {author} {\bibinfo {author} {\bibfnamefont {S.~B.}\ \bibnamefont
  {Roy}}, \bibinfo {author} {\bibfnamefont {M.~K.}\ \bibnamefont
  {Chattopadhyay}}, \bibinfo {author} {\bibfnamefont {P.}~\bibnamefont
  {Chaddah}}, \ and\ \bibinfo {author} {\bibfnamefont {A.~K.}\ \bibnamefont
  {Nigam}},\ }\href {\doibase 10.1103/PhysRevB.71.174413} {\bibfield  {journal}
  {\bibinfo  {journal} {Phys. Rev. B}\ }\textbf {\bibinfo {volume} {71}},\
  \bibinfo {pages} {174413} (\bibinfo {year} {2005})}\BibitemShut {NoStop}%
\bibitem [{\citenamefont {Givord}\ \emph {et~al.}(1983)\citenamefont {Givord},
  \citenamefont {Laforest}, \citenamefont {Lemaire},\ and\ \citenamefont
  {Lu}}]{GIV83}%
  \BibitemOpen
  \bibfield  {author} {\bibinfo {author} {\bibfnamefont {D.}~\bibnamefont
  {Givord}}, \bibinfo {author} {\bibfnamefont {J.}~\bibnamefont {Laforest}},
  \bibinfo {author} {\bibfnamefont {R.}~\bibnamefont {Lemaire}}, \ and\
  \bibinfo {author} {\bibfnamefont {Q.}~\bibnamefont {Lu}},\ }\href {\doibase
  http://dx.doi.org/10.1016/0304-8853(83)90212-3} {\bibfield  {journal}
  {\bibinfo  {journal} {J. Magn. Magn. Mater.}\ }\textbf {\bibinfo {volume}
  {31–34}},\ \bibinfo {pages} {191 } (\bibinfo {year} {1983})}\BibitemShut
  {NoStop}%
\end{thebibliography}
\end{document}